\documentclass{aa}
\usepackage{graphicx}
\usepackage{txfonts}
\usepackage{psfrag}
 
\begin{document} 
 
 \title{The dark nature of \object{GRB 051022} and its host galaxy\thanks
{Based on observations taken with the 1.0m telescope at ARIES, with the
2.0 m telescope at Hanle, with the 1.5 m Carlos S\'anchez at Observatorio 
del Teide, with the 1.5 m telescope at Observatorio de 
Sierra Nevada, with the 3.5m Telescope at the Spanish-German Calar Alto 
Observatory, with the 3.5 m Telescopio Nazionale Galileo, with the 
4.2 m William Herschel telescope, at the Observatorio del Roque de los
Muchachos in La Palma, and with the 6.0 m Bolshoi Azimuthal Telescope 
at the Special Astrophysical Observatory in Zelenchukskaya.} 
} 
 
\author{A. J.    Castro-Tirado      \inst{1} 
   \and M.       Bremer             \inst{2}
   \and S.       McBreen            \inst{3}   
   \and J.       Gorosabel          \inst{1}
   \and S.       Guziy              \inst{1,4}
   \and R. M.    Gonz\'alez Delgado \inst{1}
   \and G.       Bihain             \inst{5,6}
   \and T.       Fakthullin         \inst{7}
   \and S. B.    Pandey             \inst{1,8}
   \and M.       Jel\'{\i}nek       \inst{1}
   \and A.       de Ugarte Postigo  \inst{1}
   \and V.       Sokolov            \inst{7}
   \and K.       Misra              \inst{9}
   \and R.       Sagar              \inst{9}
   \and P.       Bama               \inst{10}
   \and A. P.    Kamble             \inst{11}
   \and G. C.    Anupama            \inst{12}
   \and J.       Licandro           \inst{4,13}
   \and F. J.    Aceituno           \inst{1}
   \and R.       Neri               \inst{2}
      }

\offprints{A.J. Castro-Tirado, \email{ajct@iaa.es} }

\institute{Instituto de Astrof\'\i sica de Andaluc\'\i a (IAA-CSIC), P.O. Box 3.004, E-18.080 Granada, Spain. 
       \and Institute de Radioastronomie Milimetrique (IRAM), 300 rue de la 
            Piscine, 38406 Saint Martin d\' \rm H\'eres, France. 
       \and Max-Planck-Institut
  f\"{u}r extraterrestrische Physik, 85748 Garching, Germany.
        \and Nikolaev State University, Nikolskaya 24, 54030 Nikolaev, Ukraine.            
       \and Instituto de Astrof\'{\i}sica de Canarias (IAC), Via L\'actea s/n,
            La Laguna, Tenerife, Spain.
       \and Consejo Superior de Investigaciones Cient\'\i ficas (CSIC), Spain.
       \and Special Astrophysical Observatory (SAO-RAS), Nizhnij Arkhyz, 
            Karachai-Cirkassian Rep., 369167 Russia 
       \and  Mullard Space Science Labratory, University College London, 
             Holmbury St. Mary, Dorking, Surrey, RH5 6NT, UK
       \and Aryabhatta Research Institute of Observational Sciences (ARIES), 
            Manora Peak, Nainital 263 129, India.
       \and Centre for Research and Education in Science and Technology 
            (CREST), Indian Institute of Astrophysics Shidlaghatta Road, 
            Hosakote 562 114, India.
       \and Raman Research Institute (RRI), Bangalore 560 080,India.
       \and Indian Institute of Astrophysics, Koramangala, Bangalore 560 034, 
            India.
       \and Isaac Newton Group of Telescopes, P.O. Box 321, E-38700 Santa 
            Cruz de la Palma (Tenerife), Spain. 
           }

\date{Received / Accepted } 
 
\abstract
{} 
  {We present multiwavelength
  (X-ray/optical/near-infrared/millimetre)  observations  of GRB 051022 between  
  2.5 hours  and $\sim$  1.15 yr  after the event. It is the  most
  intense gamma-ray burst  ($\sim$ 10$^{-4}$ erg cm$^{-2}$) detected by
  HETE-2, with the exception of the nearby GRB 030329.}
  {Optical and near infrared observations did not detect the afterglow despite 
  a strong  afterglow at X-ray wavelengths. Millimetre observations at  Plateau 
  de Bure (PdB) detected a source and a flare, confirming  the association of this event with
  a moderately bright  (R = 21.5) galaxy.}
  {Spectroscopic observations of  this galaxy show  strong [O II], H$\beta$  and [O III] 
  emission lines at a redshift of  0.809.  The spectral energy distribution
  (SED) of the galaxy implies A$_{\rm  V}$ (rest frame) =  1.0 and a
  starburst   occuring   $\sim$  25   Myr   ago,   during  which   the
  star-forming-rate reached $\geq$  25 M$_{\odot}$/yr.  In conjunction
  with the spatial extent  ($\sim$ 1$^{\prime\prime}$) it  suggests a
  very luminous  (M$_{\rm V}$  = $-$ 21.8)  blue compact  galaxy, for 
  which we also find with $Z\sim Z_\odot$.  The
  X-ray spectrum shows evidence of considerable absorption by neutral
  gas  with   N$_{\rm  H,  X-ray}$   =  3.47$^{+0.48}_{-0.47}  \times$
  10$^{22}$ cm$^{-2}$ (rest frame).
  Absorption by dust in the host galaxy at $z$ = 0.809 certainly cannot account for the 
  non-detection of the optical afterglow, unless the dust-to-gas 
  ratio is quite different than that seen in our Galaxy (i.e. large dust grains).} 
  {It is likely  that the afterglow of the dark GRB 051022 was extinguished 
  along the line of sight by an obscured, dense star forming region in
  a molecular cloud within the parent host galaxy. This galaxy is different from most 
  GRB hosts being brighter than L$^{*}$ by a factor of 3. We have also derived a 
  SFR $\sim$ 50 M$_\odot$/yr and predict that 
  this host galaxy will be detected at sub-mm wavelengths.}


\maketitle 
 
\section{Introduction} 
 
The first gamma-ray burst (GRB) with a fading X-ray afterglow 
and without an optical counterpart was
detected by {\it BeppoSAX} in Jan 1997 (Feroci et al. 1997, 
Gorosabel et al. 1998). The number 
of similar events has continued to increase 
during the Afterglow Era. 
Dark GRBs seem to constitute a significant fraction
of the GRB population (eg Filliatre et al 2006). Although 
in these cases where no optical/near infrared (nIR) afterglows have been 
detected, transient X-ray and radio emission has 
pinpointed the
parent galaxies where the bursts occurred.
About 48\% (34/71) of GRBs that were localised by {\it Swift}/XRT
during the first year of the mission 
do not show an optical (or near-IR) afterglow, 
despite deep searches 
(down to ~21-22 mag) performed for nearly all events  within 24 hr. 

Proposed explanations include 
obscuration, a low-density environment, intrinsic faintness 
and a high-redshift location 
(also discussed by Jakobsson et al. 2004, Rol et al. 
2005, Fynbo et al. 2001). 
The first possibility could be due to: i) a high column 
density of gas around the progenitor like 
a dusty, clumpy medium in a giant molecular cloud (GMCs, Lamb 2001) 
or ii) to dust in the host 
galaxy at larger distances which can account for
only $\sim$10$\%$ of the events 
(discussed in Piro et al. 2002). The occurrence of a burst in a low-density 
ambient medium (Taylor 
et al. 2000) will result in a very dim afterglow, although this 
seems unlikely due to the accepted association between
long-duration GRBs and core-collapse 
supernova whose progenitors would not have had time to
travel too far from their birth places in star-forming regions.
The high-$z$ case, in which the Ly-$\alpha$ forest emission 
will affect the optical band, it is only applicable 
for $\geq$ 10$\%$ of the events (see Gorosabel et al. 2004).

Therefore it is essential to find new potential dark GRBs and to
study whether  the {\it darkness}  is due to the  obscuration scenario
(the  a-priori   most  plausible   scenario  for  the   reasons  given
above). Thus, the bright GRB  051022 constituted a perfect case study.
It  was  discovered  by  {\it   HETE-2}  on  22  Oct  2005  (Olive  et
al.   2005).  The   burst  started   at  13:07   UT  and   lasted  for
$\approx200$~s,  putting it  in  the ``long-duration''  class of  GRBs
(Tanaka et al. 2005).  It was  also observed by {\it Mars Odyssey} and
{\it  Konus}/WIND (Hurley et  al. 2005).   It had  a fluence  of (2.20
$\pm$ 0.02) $\times$ 10$^{-5}$ erg cm$^{-2}$ in the 2-30 keV range and
(1.40 $\pm$ 0.02)  $\times$ 10$^{-4}$ erg cm$^{-2}$ in  the 30-400 keV
energy  band, making  it the  highest fluence  event detected  by {\it
HETE-2} in  its 6-yr  lifetime, with the sole exception of the  nearby GRB
030329 (Ricker 2005).  The average spectrum of the prompt emission was
best   fit  by   an  absorbed   Band  model   with  $\alpha$   =  1.01
$^{+0.02}_{-0.03}$,                     $\beta$=1.95$^{+0.25}_{-0.14}$,
E$_{peak}$=213$\pm18$            keV            with           N$_{\rm
H}$=(1.51$^{+0.53}_{-0.50}) \times$10$^{22}$ atoms cm$^{-2}$ (Nakagawa
et al. 2006).
Golenetskii et al. (2005) report the best fit to 
the Konus data is a power law with exponential cutoff model with
alpha = 1.176 $\pm$0.038  and  E$_{peak}$ = 510 $^{+37}_{-33}$) keV 
in the 20 keV - 2 MeV energy range, a higher value than derived 
by \textit{HETE-2}.   
The prompt dissemination (55.0 s) of the GRB position by {\it HETE-2}  
enabled rapid responses by automated and robotic telescopes 
like ROTSE-III (Schaefer et al. 2005) and ART (Torii et al. 2005), 
although no prompt optical afterglow was detected.
{\it Swift} slewed and started data acquisition (about 3.5 
hr after the event) and fading X-ray emission was detected by the 
XRT, which can be considered as the first clear detection of 
the afterglow from GRB051022 (Racusin et al. 2005a). This result 
triggered a 
multiwavelength campaign at many observatories aimed at detecting the 
afterglow at other wavelengths.

Here we report the results of the multiwavelength observations carried out,
from millimetre wavelengths to the X-ray band, for both the afterglow and 
for the host galaxy, 
and we discuss  the implications of the study for dark GRBs.

\begin{figure} 
\begin{center}
      \resizebox{8.5cm}{!}{\includegraphics{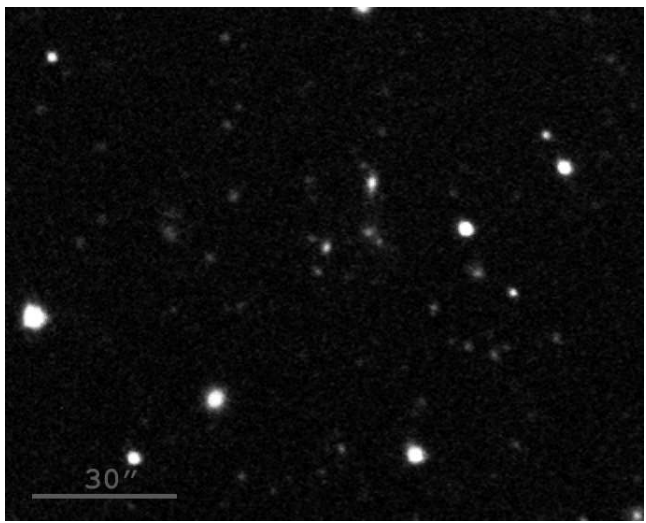}} 
      \caption{The $BVR$-band composite image of the \object{GRB 051022} 
               field  
               taken at the 1.5m OSN on 26 Oct 2005. The potential host galaxy
               is the bluest object in the field, close to the centre of 
               the image. The field is 
               $2^{\prime}.1 \times 1^{\prime}$.7 with 
               North up and East to the left.} 
    \label{carta ID}
\end{center} 
\end{figure} 
 
\begin{table*} 
      \begin{center} 
            \caption{Journal of optical and near-infrared (nIR) frames taken on 
the \object{GRB 051022} field.} 
                     \begin{tabular}{@{}lcccc@{}} 
 
Date of exposure UT & Telescope/ & Filtre/ & Exposure Time & Limiting \\ 
(mid exposure)  & Instrument & Grism   &    (seconds)  & Magnitude \\  
\hline 
22 Oct 05, 16:18   & 2.0HCT (HFOSC)     & $R$  &      600   &  22.5   \\
22 Oct 05, 17:00   & 2.0HCT (HFOSC)     & $I$  &      600   &  21.8   \\
22 Oct 05, 17:20   & 1.0ST (CCD)        & $R$  &      900   &  20.8   \\
24 Oct 05, 20:45   & 1.5OSN (CCD)       & $B$  &   1\,800   &  23.0   \\
24 Oct 05, 22:15   & 1.5OSN (CCD)       & $V$  &      720   &  22.5   \\
24 Oct 05, 19:40   & 1.5OSN (CCD)       & $R$  &      120   &  22.2   \\
24 Oct 05, 19:43   & 1.5OSN (CCD)       & $I$  &      120   &  21.3   \\
26 Oct 05, 20:00   & 1.5OSN (CCD)       & $U$  &   3\,000   &  22.9   \\
31 Oct 05, 03:30   & 1.5OSN (CCD)       & $U$  &   6\,000   &  23.4   \\
07 Nov 05,  22:47   & 4.2WHT (ISIS)      & R300B &  5\,400   &   --    \\
07 Nov 05,  22:47   & 4.2WHT (ISIS)      & R300B &  1\,800   &   --    \\
07 Nov 05,  23:24   & 4.2WHT (ISIS)      & R300R &  3\,600   &   --    \\
12 Dec 06, 17:02 & 6.0BTA (SCORPIO)     & VPHG400 &  7\,200 &   --     \\
\hline
23 Oct 05, 01:56   & 1.5TCS (CAIN-2)    & $K_{S}$ &   1\,800   &  18.0   \\
23 Oct 05, 02:42   & 1.5TCS (CAIN-2)    & $J$  &   1\,800   &  19.0   \\
23 Oct 05, 23:30   & 1.5TCS (CAIN-2)    & $H$  &   1\,800   &  18.5   \\
24 Oct 05, 00:20   & 1.5TCS (CAIN-2)    & $K_{S}$ &   1\,800   &  18.0   \\
20 Nov 05, 20:52	& 3.5TNG (NICS)	     & $J$  &   1\,200	 &  20.0   \\
20 Nov 05, 21:28	& 3.5TNG (NICS)	     & $H$  &   1\,200	 &  19.5   \\ 
14 Oct 06, 00:39 & 3.5CAHA (OMEGA2000)	     & $H$  &   2\,400	 &  20.0   \\
\hline 
                         \label{tabla1} 
                     \end{tabular} 
      \end{center} 
\end{table*}

\begin{table*} 
      \begin{center} 
            \caption{Journal of millimetre (mm) observations of the GRB 051022 field.} 
                     \begin{tabular}{@{}lccccc@{}} 
 
Date of 2005 UT             & Configuration & Flux             & Frequency & Beam   & Position angle \\ 
(start time - end time)              &               & (mJy, 1$\sigma$) &   (GHz)   & (arcsec) & ($^{\circ}$) \\  
\hline 
23.Oct 20:42 - 23.Oct 23:27 & 6Dp &  1.14 $\pm$ 0.28 &  90.808 &  8.7  x 4.0  &  -103 \\
                            &     &  0.02 $\pm$ 1.68 & 217.827 &  8.9  x 1.7  &  + 64 \\
24.Oct 19:17 - 25.Oct 01:44 & 6Dp &  0.08 $\pm$ 0.17 &  86.243 &  9.1  x 4.6  &  + 79 \\
                            &     & -2.42 $\pm$ 0.85 & 232.032 &  4.7  x 1.6  &  + 73 \\
28.Oct 20:44 - 29.Oct 01:28 & 6Dp & -0.33 $\pm$ 0.44 &  86.243 &  9.2  x 3.8  &  - 96 \\
                            &     &  1.81 $\pm$ 3.40 & 232.032 &  2.5  x 1.8  &  + 57 \\
30.Oct 15:27 - 30.Oct 18:20 & 5Dp &  3.48 $\pm$ 1.05 &  90.229 &  10.7 x 3.5  &  - 57 \\
01.Nov 20:44 - 01.Nov 01:28 & 6Dp & -0.50 $\pm$ 0.26 &  86.243 &  8.0  x 3.8  &  - 69 \\
                            &     &  0.55 $\pm$ 1.35 & 221.501 &  3.0  x 1.4  &  - 63 \\
\hline 
                         \label{tabla2} 
                     \end{tabular} 
      \end{center} 
\end{table*}

\begin{table*} 
      \begin{center} 
            \caption{GRB 051022 host galaxy optical and nIR photometry.} 
                     \begin{tabular}{@{}cccccccc@{}} 
 
       U        &        B       &        V       &        R       &     I          &      J         &         H      &  K$_{\rm \ S}$  \\ 
                &                &                &                &                &                &                &                \\ 
\hline 
22.60 $\pm$ 0.20&22.61 $\pm$ 0.10&21.93 $\pm$ 0.12&21.50 $\pm$ 0.10&20.84 $\pm$ 0.12&20.36 $\pm$ 0.12&19.30 $\pm$ 0.08&18.15 $\pm$ 0.23\\
\hline 
                         \label{tabla3} 
                     \end{tabular} 
      \end{center} 
\end{table*}

\section{Observations and data reduction} 
  \label{observaciones}

  \subsection{Optical and near-IR observations}   
  Target of opportunity (ToO) observations 
  in the optical were started  
  2.9~hr after the event using the
  1.0~m telescope (+CCD) at ARIES and  
  the 2.0~m telescope (+ HFOSC 
  camera) at Hanle (near Himalaya in India). Additional optical observations  
  were made at 1.5~m telescope (+ CCD) at
  Observatorio de Sierra Nevada in Granada (Spain).
  Near-IR (nIR) observations were started at the 
  1.5~m Carlos S\'anchez telescope 
  (+ CAIN-2) at Observatorio the Iza\~na en Tenerife (Spain) and with the 
  3.5~m Telescopio Nazionale Galileo (+ NICS) at Observatorio del Roque de Los 
  Muchachos in La Palma (Spain). Spectroscopic observations of 
  the potential host galaxy were conducted at the 4.2m William Herschel 
  Telescope at La Palma on 7 
  Nov 5, using ISIS with the blue arm centred at 4500 $\rm \AA$ 
  and the R300B grating (3200-5300 $\rm \AA$, dispersion 0.86 $\rm \AA$/pix) 
  and the red arm centred at 6450 $\rm \AA$ and the R300R grating 
  (5300-7600 $\rm \AA$, dispersion 0.84 $\rm \AA$/pix). A second 
  spectrum was obtained with the red arm centred at 8650 $\rm \AA$ and
  the R300R grating (7500-9800  $\rm \AA$, dispersion 0.86 $\rm \AA$/pix). 
  A third spectrum was obtained at the 6.0m Bolshoi Azimuthal Telescope 
  in Caucasus on 12 Dec 2006, using SCORPIO with the VPHG400 holographic grism 
  (3500-9500  $\rm \AA$, 15 $\AA$ resolution).
  The observation log is presented in Table \ref{tabla1}.

  In order to determine the magnitudes in all our optical and nIR images, we 
  used aperture photometry with the DAOPHOT routine under 
  IRAF\footnote{IRAF  is   distributed   by the   NOAO, 
  which are operated by USRA, 
  under  cooperative  agreement with  the US  NSF.}.  
  The potential host galaxy was revealed in the optical (Fig. 1). 
  The optical field was calibrated  using the calibration provided by 
  Henden (2005).
  The nIR images were calibrated using the 2MASS Catalogue.
  
  The spectroscopic observations were reduced under IRAF following the 
  standard procedure to correct the frames from bias and flatfield. The 
  spectra are also calibrated in wavelength, but not in flux.
 
  \subsection{Millimetre observations}
  Additional mm observations were obtained at the Plateau de Bure 
  Interferometer (PdBI) as part of an on-going ToO programme. 
  The PdBI observed the source on Oct 23, 24, 28 and 31, in the 5 antenna 
  compact D configuration).
  The data reduction was done with the standard CLIC and MAPPING software 
  distributed by the Grenoble GILDAS group, the flux calibration is relative 
  to the carbon star MWC349. Amplitude and phase calibration were relative 
  to the quasar 3C454.3.  
  The observing log is presented in Table \ref{tabla2}. 
 
  \subsection{X-ray observations}
  The XRT on {\it Swift} began to observe GRB 051022 $\sim$3.5 hours after the 
  trigger and detected a decaying X-ray source in 
  the {\it Swift}/XRT field at position 
  RA(J2000)=23$^{\rm h}$56$^{\rm m}$4.1$^{\rm s}$, 
  Dec(J2000)=+19$^\circ$36$^\prime$25$\prime\prime$ 
  (Racusin et al. 2005a).
  The XRT observations are in  Photon 
  Counting (PC) mode and were reduced using the standard xrtpipeline 
  (version 0.10.4) for XRT data analysis 
  software\footnote{http://swift.gsfc.nasa.gov
  /docs/software/lheasoft/download.html}   using the most recent 
  calibration files. The spectral data were analysed with the 
  XSPEC version 11.3 (Arnaud 1996).

  The  first two observations  show evidence  of mild  pile up  and an
  annular source region rather than a circular one was used in both of
  these cases  (for a  detailed discussion  of PC pile  up in  XRT see
  Vaughan et al. 2006).  The  radius of the affected inner annulus was
  determined  by fitting  the point  spread function  to the  data and
  selecting  regions where  the data  are are  well fit  by  the point
  spread  function.  Inner  annuli regions  of  10 and  5 arcsec  were
  excluded in  this manner for orbit  1 and 2  respectively. The
  exposure map correction was used  was to account for the presence of
  bad columns close  to the centre of the  source image.  In addition,
  the  count  rate was  manually  corrected  for  the loss  of  counts
  incurred by the use of an  annular extraction region for the pile up
  correction.
 
 \section{Results and discussion} 
  \label{resultados} 
  The afterglow was detected at X-ray and millimetre wavelengths, but not 
  in the optical of near infrared bandpasses.  
  The later non-detections are in agreement with the upper 
  limits reported by 
  ROTSE-III (Schaefer et al. 2005) and ART (Torii et al. 2005).

    \subsection{No optical/nIR afterglow } 
             \label{sin emision} 
              
The {\it Swift}/XRT detected the X-ray afterglow 
for \object{GRB 051022} (Racusin et al. 2005a), allowed us to promptly 
identify two objects (dubbed ``A'' and ``B'', adopting the naming convention 
introduced by Castro-Tirado et al. 2005) in the 
surroundings of the
 XRT X-ray error box. Only object ``B'' was located inside the box
while "A'' was just 1.5 arcsec outside. Neither of the two objects were found 
to be varying in our images, in agreement with other early time reports by 
Nysewander et al. (2005) or Burenin et al. (2005). 

In order  to estimate the optical and  nIR limiting magnitudes
of  the  possible  underlying   transient  at  these  wavelengths,  we
simulated a point source underlying  in the host galaxy, making use of
the PSF in nearby stars in the  field in both the deep $R_{\rm c}$ and
$K_{\rm  S}$-band images,  and  determined the  limiting magnitude  at
which  the simulated  point source  would  be remained  detected at  a
3$\sigma$ level  by our detection  software. 
These values, $R$  = 21.5 and $K_{\rm S}$ = 18.0, are used hereafter.

   \subsection{Millimetre afterglow: flaring activity ?} 
             \label{flaring} 

	In order to detect the elusive afterglow, mm observations where 
conducted at PdBI 33 hr after the burst. 
The afterglow was successfully detected, 
with a flux density of 1.14 $\pm$ 0.28 mJy at 
90 GHz, at a position 
 RA(J2000)=23$^{\rm h}$56$^{\rm m}$04.15$^{\rm s}$, 
  Dec(J2000)=+19$^\circ$36$^\prime$25\farcs1 
($\pm$ 0$^{\prime\prime}$.6). The source was
coincident, within errors, with the above-mentioned source B, which was found 
to be extended on images taken by Berger and Wyatt (2005).
Thus, the position of the millimetre afterglow position is superimposed 
  on a bright optical/nIR potential host galaxy. 
The mm afterglow faded substantially in 24 hours
and the object was not detected 
 on 24 Oct or 28 October. Another detection was obtained on 31 October
(see Fig. 2), thus confirming the trend observed at 4.8 GHz 
(Fig. 1 of Rol et al. 2007).
Such a mm reflaring at late epoch might be caused
by an energy injection episode 
but the lack of detailed observations at other wavelengths 
prevents confirmation.
 
We also unsuccessfully searched for line emission around the 3mm wavelength 
and therefore ``cut'' the spectral information of the 90 GHz observations into 
large velocity bins (100, 200 km s$^{-1}$) to check if there is a line profile 
(which should have a signal-to-noise ratio $\geq$ 3, because the average over 
the whole continuum band would mix the line with noise and weaken it).

\begin{figure}[t] 
\begin{center}
      \resizebox{7.5cm}{5.9cm}{\includegraphics{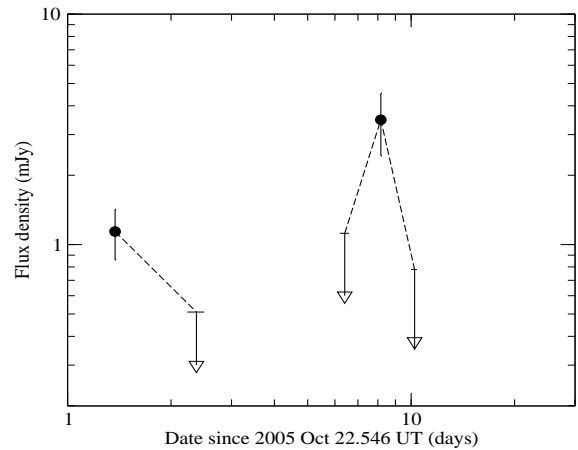}} 
      \caption{The 3 mm wavelength afterglow lightcurve of GRB 051022 obtained 
       at PdBI. Detections were obtained at 90 GHz (filled circles) whereas 
       only upper limits were obtained at 86 GHz.} 
    \label{carta ID}

\end{center} 
\end{figure}

        \subsection{The X-ray afterglow} 
        \label{x-ray}         
        
\begin{figure} 
\begin{center}
   \resizebox{8.5cm}{!}{\includegraphics{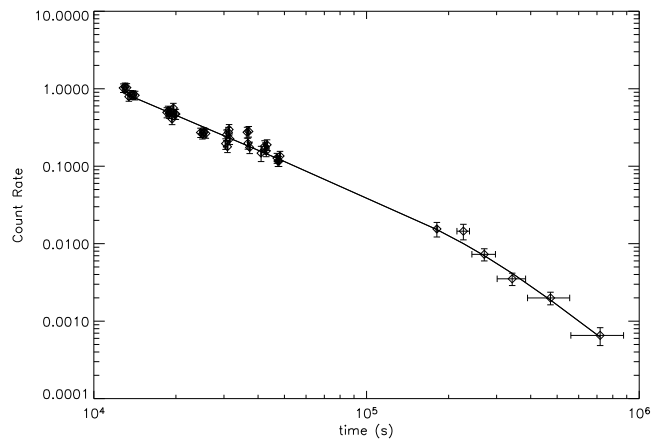}} 
      \caption{The {\it Swift}/XRT light curve of GRB~051022. A break at 
       $\sim$ 2 $\times$ 10$^{5}$ s is apparent on the data.} 
    \label{xraylc}
\end{center} 
\end{figure} 

\begin{figure} 
\begin{center}
   \resizebox{8.5cm}{!}{\includegraphics{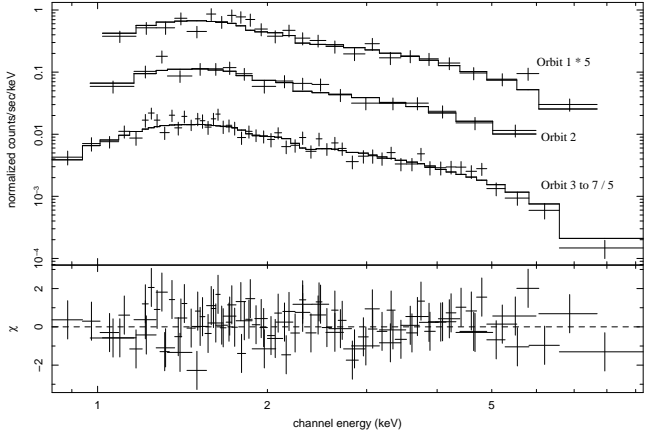}}
      \caption{Pre-break spectra of GRB~051022 including orbits 1, 2 and 3 to 7 separately. The spectra are offset as indicated for clarity of presentation.  } 
    \label{xrayspec}
\end{center} 
\end{figure}

The X-ray light curve of GRB~051022 in the energy range 0.3 to 10 ~keV
is shown in Fig.~\ref{xraylc}. The data  is best fit by a broken power
law  with pre-break  index,  $\alpha_{X1}$= 1.50$\pm$0.07,  post-break
index $\alpha_{X2}$=2.47$\pm$0.40 with  a break at 216$\pm$55~ks using
a  smoothing factor of  10.  The  fit is  roughly consistent  with the
values reported by Racusin et al (2005b) who obtain a pre-break power law index of 1.33$\pm$0.07, a  
break at 2.9 $\pm$ 0.2 days (250$\pm$17~ks)  
after which the power-law index changed to 3.6$\pm$0.4 and the differences 
can likely be attributed to the availability of new software and responses.
Rol et al (2007) report an earlier break time of  
110$^{+21}_{-23}$ks based on the \textit{Swift}/XRT
and \textit{Chandra} data.

The spectra shown in Fig.~\ref{xrayspec} were extracted individually for 
orbit 1, 2 and 3 to 7 and were fit simultaneously.  The data have 20 counts 
per bin. The pre-break X-ray spectra  can be fit by a photon index $\Gamma$ = 
2.25 $\pm$ 0.15 and a column density of 
(0.85$^{+0.12}_{-0.11}$) $\times$ 10$^{22}$ cm$^{-2}$ 
with $\chi^{2}/dof$ = 81/93.  A value of 
3.47$^{+0.48}_{-0.47}$ $\times$ 10$^{22}$ cm$^{-2}$ was obtained for the 
extra absorption by a cold neutral gas within the host galaxy at  redshift 
0.809, in addition to the absorption within our own galaxy which is 
4.09 $\times$ 10$^{20}$ cm$^{-2}$ (Dickey \& Lockman 1990) .  
We find no evidence for spectral evolution in the X-ray spectrum of 
GRB~051022. 
This analysis is in agreement with the XRT results obtained by Butler et 
al. (2005) and and Rol et al. (2007). 
Nakagawa et al. (2006) derive a higher value of the column density 
of 8.8$^{+1.9}_{-1.8} \times 10^{22}$  cm$^{-2}$ (rest frame) in the prompt emission, 
and Rol et al. (2007) point out that this could indicate a change in absorption
between the early HETE observations and those of the XRT.

\subsection{Spectral Energy Distribution modelling} 
        \label{OA sed}


        We  have  determined the  spectral  energy distribution  (SED)
        for \object{GRB 051022} at T$_{0}$ + 33 hr at time of the
        PdB observations (Fig 5.).  The  data are well fit by  a jet model with
        $p$ = 2.6 $\pm$ 0.2, as  $\alpha$ = (3$p$ $-$ 2)/4 and $\beta$
        =  $p$/2  for pre-break  (isotropic  emission)  and $\nu_{\rm X}$  
        $>$ $\nu_{c}$  derived from  the
        X-ray data  and assuming the slow-cooling  regime (Sari  et  al. (1998)).  We cannot
        distinguish  whether  the external  medium  is  ISM ($\rho$  =
        constant)  or   a  stellar  wind   profile  ($\rho$  $\propto$
        $r^{-2}$).  We have also  included the radio detections at the
        WSRT and  VLA (van  der Horst et  al. 2005, Cameron  and Frail
        2005).

\subsection{The obscuring medium} 
        \label{obscuration}

	GRB 051022  is located
        undoubtely  in the  dark GRB  locus of  the F$_{\rm  opt}$ $-$
        F$_{\rm X-ray}$ diagramme of Jakobsson et al. (2004).  Clearly
        the reason that the afterglow was not detected either at optical 
        or at nIR wavelengths is due to obscuration along the
        line of sight.
        The spectral energy distribution can be used to 
        estimate the amount of extinction by dust at optical 
        wavelengths and 
        predicts a magnitude $R$ $\sim$ 18.3 at 
        T$_{0}$ + 33 hr. 
        The $R$ band upper limit from our observations is 21.5 mag and therefore
        we infer a considerable extinction 
        toward GRB 051022, with a lower limit A$_{\rm R}$ = 3.2 mag in the
        observer frame.
        This is equivalent to A$_{\rm V}$ = 1.7 mag in the 
        rest frame at $z$ = 0.809 (see next subsection) from which an 
        equivalent Hydrogen column density at rest frame of 
        N$_{\rm H, opt}$ $\geq$ 0.4 $\times$ 10$^{22}$
        cm$^{-2}$ is derived, assuming the relationship 
        A$_V$ = 0.56 $\times$ N$_{\rm H}$ (10$^{21}$ cm$^{-2}$) 
        + 0.23 of Predehl \& Schmitt (1995).   
        A value of 
	N$_{\rm H, X-ray}$ = (3.47 $^{+0.48}_{-0.47}$) 
        $\times$ 10$^{22}$ cm$^{-2}$  
        (rest frame) was derived from the X-ray data assuming the 
        absorbing material being in a neutral cold state. 
        Therefore, the ratio of  N$_{\rm H, X-ray}$/N$_{\rm H, opt}$ 
        less than 9.
 
        Comparable values of N$_{\rm H, X-ray}$ (rest frame) 
        were found in other GRBs like \object{GRB 980703} 
        (Galama \& Wijers 2001) and \object{GRB 050401} 
        (Campana et al. 2006) and whereas substantially higher 
        values were found in another two dark GRB host galaxies 
        \object{GRB 970828} (Djorgovski et al. 2001) and \object{GRB 000210} 
        (Piro et al. 2002), assuming  the 
        gas-to-dust ratio is compatible with the Galactic value.

        Is it possible that 
        the absorption took place in the circumburst environment
        of the GRB  by the  ISM  of the  host galaxy? The value  of
        A$_{\rm  V}$ =  1.0  derived  from the  host  galaxy SED  (see
        subsection  3.4) cannot  account  for the  properties of  this
        event assuming a dust-to-gas  ratio similar to the Galactic
        value.  Therefore the remaining possibility is that
        absorbing medium is a  giant molecular cloud (GMC) along
        the  line of  sight and within  the parent  host galaxy
        where such high column densities would be expected. 
        This is also supported
        by the high absorption column densities (within errors) 
        derived from the {\it 
        HETE-2} and {\it Swift}/XRT observations 
        (see also Nakagawa et al. 2006).

\subsection{The nature of the host galaxy} 
        \label{host spectroscopy} 
	
	The optical  spectrum of the  GRB 051022 host galaxy  (Fig. 6)
        shows one prominent  emission line at 6731 $\rm  \AA$ which we
        interpreted as [O II] 3727 $\rm \AA$ at a redshift $z$ = 0.806
        $\pm$ 0.001.  This is confirmed by the additional detection of
        [O III] 5007 $\rm \AA$ at  9071 $\rm \AA$ and H$\beta$ at 8806
        $\rm \AA$ consistent with $z$ = 0.811.  
        The [O III]  4946 $\rm \AA$ falls within prominent
        sky  lines and  could not  be accurately determined.    
        The emission lines  strengths were measured  using a  Gaussian fit  
        in IRAF $splot$.  No correction  for underlying  Balmer  absorption is
        applied. 
        The details are  presented  in Table 4.  
        This  redshift is in
        agreement with the value $z$ = 0.8 reported by Gal-Yam et
        al. (2005). The  corresponding  luminosity distance  is
        D$_{L}$  = 5.09  $\pm$ 0.20  Gpc (assuming  H$_{0}$ =  71 km
        s$^{-1}$     Mpc$^{-1}$,   $\Omega_{m}$  =  0.27  and
        $\Omega_{\Lambda}$ = 0.73),
        from which Rol et al. (2007) derived a collimation-corrected 
        energy released in gamma-rays for GRB 051022 of 
        (8-18) $\times$ 10$^{50}$ erg.
        
        We derive  a lower limit on the  star formation rate (SFR)
        from the strength of the [O II] line, applying the calibration
        of Kennicutt (1998), 
        SFR  (M$_{\odot}$/yr) = (1.4  $\pm$ 0.4) $\times$  10$^{-41}$  
        L$_{[O II]}$.     Using the 
        measured   line    luminosity   as   a    lower   limit implies  SFR
        (M$_{\odot}$/yr)  $>$  3.2  M$_{\odot}$/yr and the SFR becomes
        SFR (M$_{\odot}$/yr) = (48 $\pm$ 3) 
        M$_{\odot}$/yr when the intrisic A$_{V}$ = 1.0 is taken into 
        account.   
       
        Another independent  measurement for  the SFR can  be obtained
        from the UV continuum at  2800 {\AA}, which is produced by massive
        stars in the galaxy.   Using the host galaxy SED, and the 
        Kennicutt (1998) estimator, (M$_{\odot}$/yr)  =  1.4 $\times$  
        10$^{-28}$ L$_{\nu(UV)}$ (erg s$^{-1}$ Hz$^{-1}$), 
        we derive a lower limit for the SFR(UV)(M$_{\odot}$/yr) $\geq$ 8 
        M$_{\odot}$/yr, which becomes SFR(UV)(M$_{\odot}$/yr) = 59 
        M$_{\odot}$/yr when the intrisic A$_{V}$ = 1.0 is taken into 
        account.

       The detection of  $[$O II$]$ {\rm $\lambda$3727}, H$\beta$
       {\rm $\lambda$4861} and $[$O III$]$ {\rm $\lambda$5007} allowed
       us to estimate the  value of the $R_{23}$ metallicity indicator
       (Pagel et al.  1979).  To  calculate $R_{23}$ we assumed a flux
       ratio of 1/3 of $[$O  III$]$ {\rm $\lambda$4959} with respect to
       $[$O  III$]$  {\rm  $\lambda$5007}.   Assuming  a  host  galaxy
       extinction of  $A_{\rm V}=1.0$  mag, a SMC-like  extinction law
       (Pei  1992),   and  correcting  for   the  Galactic  extinction
       ($E(B-V)=0.06$, according to Schlegel et al.  1998) we obtained
       a  value of  $R_{23}=4.0\pm0.3$.  However, without  detections  of other
       emission lines (like $[$N II$]$)  we can not break the $R_{23}$
       degeneracy, and there are two possible metallicity values. Given
       that our $[$O III$]$/ $[$O  II$]$ flux ratio (once is corrected
       for  Galactic and  host  galaxy extinction)  is  $\sim 1$,  our
       $R_{23}$ yields 12+log(O/H) values of $\sim 7.7$ (lower branch)
       and  $\sim 8.7$ (upper  branch), respectively  (see Fig.   4 of
       Maier  et al.  2004).   The metallicity-luminosity  relation at
       medium  redshift  galaxies   ($z=0.64$,  Maier  et  al.   2004)
       predicts a  metallicity around  12+log(O/H) $\sim 9.1$  for the
       bright absolute magnitude of  our host (M$_{B}$ = -21.8), hence
       supporting  the metallicity  obtained with  the  upper R$_{23}$
       branch consistent with $Z\sim Z_\odot$.

\begin{table*} 
      \begin{center} 
      \caption{Emission lines detected in the \object{GRB 051022} host galaxy.} 
                     \begin{tabular}{@{}lcccc@{}} 
 
Line ID                         &  $\lambda_{obs}$  &  $z$  & Flux (10$^{-16}$ erg cm$^{-2}$ s$^{-1}$)\\ 
\hline 
$[$O II$]$  \rm $\lambda$3727   &  6730.7  &  0.806  &  9.4$\pm$0.3  \\
H$\beta$    \rm $\lambda$4861   &  8806.3  &  0.811  &  7.9$\pm$0.6  \\
$[$O III$]$ \rm $\lambda$5007   &  9070.6  &  0.811  & 13.4$\pm$0.5  \\
\hline 
                         \label{tabla4} 
                     \end{tabular} 
      \end{center} 
\end{table*}

	The spectral energy distribution of the GRB 051022 host galaxy
        is shown in Fig. 7.   The restframe colours (and therefore the
        associated  K-corrections)  have been  obtained  based on  the
        HyperZ code  (Bolzonella et  al. 2000), fitting  synthetic SED
        templates  to our  $UBVRIJK_{S}$-band magnitudes  derived from
        our own data (see Table  3 and correcting all for the Galactic
        reddening  E(B-V)= 0.06,  Schlegel  et al.  1998).   At $z$  =
        0.809, this is a rather  luminous galaxy, with M$_{B}$ = -21.8
        (or  3 $\times$  L$^{\star}$ adopting  M$_{B}$ =  -20.6  for a
        L$^{\star}$ galaxy from Schechter (1976)). 

        Using an apparent size of $\sim$ 1$^{\prime\prime}$
        in our images, 
        this would correspond to 
        about 8 kpc at $z$ = 0.809, typical of compact galaxies and 
        therefore it is possible that the GRB 051022 host would be 
        a blue compact galaxy. Its M$_{B}$ value would place it at
        the top of their luminosity distribution (Bergvall \& 
        \"Ostlin 2002).

        The mass of the GRB 051022 can be determined using several
        methods: i) from the fitting performed with HyperZ, we get
        a lower limit of $>$4.3 $\times$ 10$^{9}$ M$_{\odot}$; ii) 
        following Bell et al (2005), a value of 
        2 $\times$ 10$^{10}$ M$_{\odot}$ is derived;
        and iii) 4 $\times$ 10$^{11}$ M$_{\odot}$ from Brinchmann 
        \& Ellis (2000). The large scatter 
        is caused by the asumptions inherent to each method.

        How does the GRB 051022 host compare to other dark GRB host 
        galaxies ?         
        Table 5 displays several properties for three dark/grey GRB host 
        galaxies:
        \object{GRB 000210} (Gorosabel et al. 2003a), \object{GRB 000418} 
        (Gorosabel et al. 2003b) and \object{GRB 051022}. 
        As it can be seen, the GRB 051022 host galaxy seems to be a luminous 
        galaxy (M$_{B}$ = $-$21.8), with a SFR of $\sim$ 50 M$\odot$/yr$^{-1}$, 
        among the highest SFR found in GRB host galaxies.

 \begin{figure}
 \begin{center} 
 \resizebox{8.5cm}{6.5cm}{\includegraphics{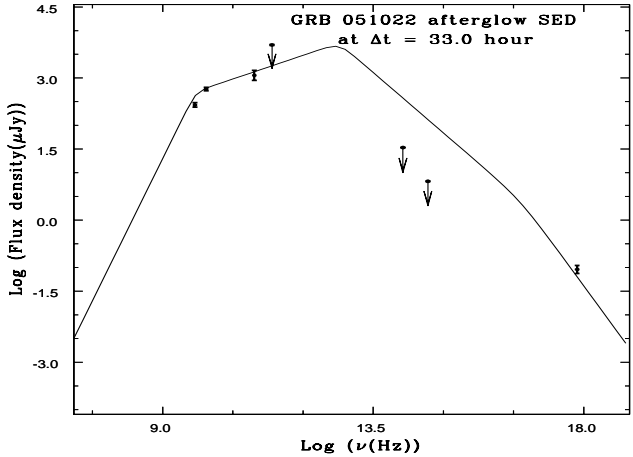}} 
 
 \caption{ The  SED of  the GRB  051022 afterglow at  T$_{0}$ +  33 hr
         (prior  to the  2.9 day  break  time reported  by Racusin  et
         al.   (2005)).   We   consider  $p$   =  2.6,   assuming  the
         slow-cooling  regime.   We  have  also   included  the  radio
         detections at  the WSRT and VLA  (van der Horst  et al. 2005,
         Cameron and Frail  2005), as well as the  220 GHz and optical
         and near-IR upper limits  reported in this paper. The initial
         parameters given at  T$_{0}$ + 0.06 days are  $\nu_{a}$ = 5.1
         $\times$ 10$^{9}$ Hz, $\nu_{m}$  = 6.5 $\times$ 10$^{14}$ Hz,
         $\nu_{c}$ =  2.1 $\times$ 10$^{17}$  Hz and F$_{\nu,  max}$ =
         5.5 mJy.}
 
    \label{OA sed}
  \end{center} 
\end{figure}

\begin{figure}
 \begin{center} 
 \resizebox{9.5cm}{!}{\rotatebox{-90}{\includegraphics{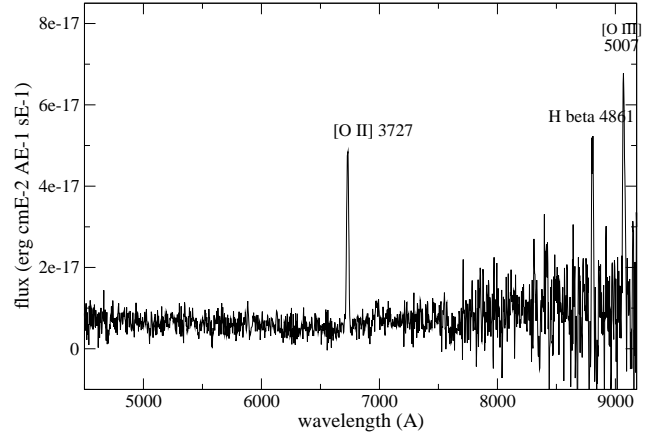}}} 
 
 \caption{     
        The optical spectrum of the GRB 051022 host galaxy taken at the 
        6.0m BTA (+SCORPIO) on 12 Dec 2006. Residuals of sky emission features
        are visible above 7700 {\AA}. The three emission lines 
        detected are consistent with z = 0.809 $\pm$ 0.002.}  
 
    \label{spectrum}
  \end{center} 
\end{figure}

\begin{figure}
 \begin{center} 
 \resizebox{9.5cm}{!}{\rotatebox{-90}{\includegraphics{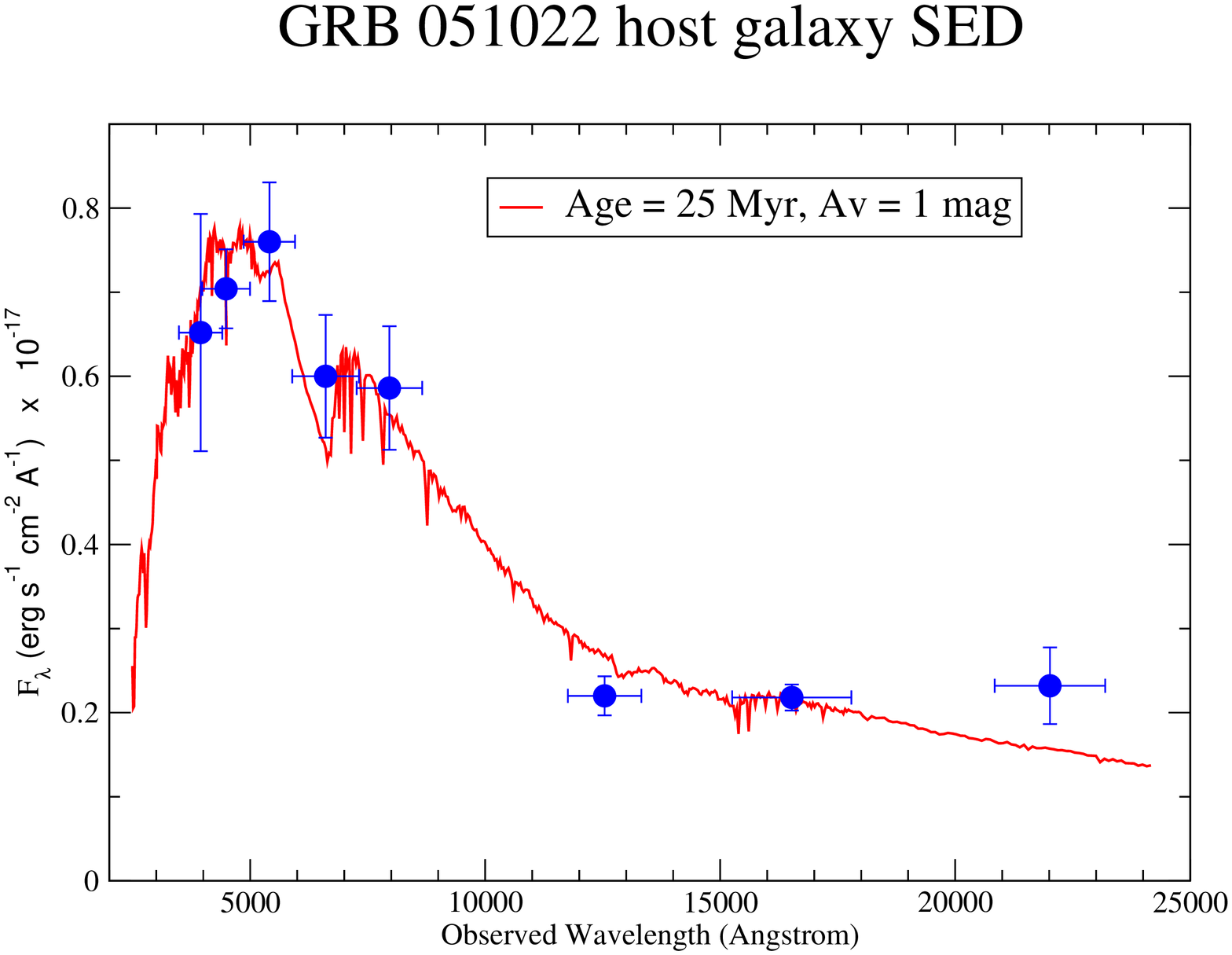}}} 
 \caption{ The  SED of the  GRB 051022 host  galaxy from the U  to the
        K$_{S}$ band.   The data are well fit  ($\chi^{2}/dof$ = 1.04)
        by a Im model with considerable extinction, A$_{\rm V}$ (host)
        = 1.0  mag and by a  starburst episode starting  $\sim$ 25 Myr
        ago.}
 
    \label{host sed}
  \end{center} 
\end{figure}

\begin{table*} 
      \begin{center} 
            \caption{Host galaxies of dark/grey GRBs.} 
                     \begin{tabular}{@{}lccccc@{}} 
 
GRB      & redshift & A$_{V}$ (host) & M$_{B}$  & Starburst age & SFR              \\ 
(YYMMDD) &          &  (mag)         &          &    (Myr)      & (M$_{\odot}$/yr) \\  
\hline 
000210   & 0.842    &       0        &  $-$19.0 &  $\sim$180    &       2.1        \\
000418   & 1.118    &      0.12      &  $-$20.6 &  $\sim$50     &       7.6        \\
051022   & 0.809    &      1.0       &  $-$21.8 &  $\sim$25     &      $\sim$50    \\
\hline 
                         \label{tabla5} 
                     \end{tabular} 
      \end{center} 
\end{table*}

\section{Conclusions} 
  \label{conclusiones} 
   
  We have shown multiwavelength observations of the long duration  
  gamma-ray burst detected by {\it HETE-2} (GRB 051022) between 2.5   
  hours and $\sim$ 30 days after the  
  event. Although a mm transient was found, no optical/nIR afterglow emission 
  has been detected at the position of the X--ray afterglow detected by 
  {\it Swift}.
  
  The X-ray spectrum show evidence of absorption by neutral gas with 
  N$_{\rm H, X-ray}$ = 0.85 $\times$ 10$^{22}$ cm$^{-2}$ (in the 
  observer frame), but ISM absorption by 
  dust in the host galaxy at $z$ = 0.809 cannot fully account for the 
  non-detection of the optical (and nIR) afterglow. 
  Therefore we are left with the possibility that that the 
  afterglow was extinguished along the line of sight by
  a giant Molecular cloud (GMC) where the N$_{\rm H, X-ray}$ values 
  are typical to the one
  found here and therefore no optical/nIR afterglow could be detected as 
  A$_{\rm V}$ $\sim$ 50 mag (A$_{\rm K}$ $\sim$ 6 mag).

  GRB 051022 is one of the most intense dark GRBs detected in the afterglow 
  era and its host galaxy is, unlike, most of the GRB hosts, brighter than 
  L$^{*}$. We have also shown that the host galaxy is forming stars at a 
  significant rate and derived a SFR (UV) $\geq$ 8 M$_\odot$/yr. We foresee 
  that this galaxy will be detected as sub-mm wavelengths as was the case 
  for the host galaxies 
  of GRB 000210 and GRB 000418 (Berger et al. 2003).

  The synergy between missions like {\it HETE-2} and {\it Swift}, 
  the latter being 
  able to be repointed and detect the X-ray afterglow, will allow to study the 
  population of dark GRBs to determine if extinction in the 
  host
  galaxy (as we have argued in the case for GRB 051022) is the reason 
  that about one-half of afterglows are beyond the 
  reach of  current optical telescopes.

\begin{acknowledgements} 
   
  We thank the anonymous referee for useful comments. 
  This work is based partly on
  observations carried out with the IRAM Plateau de Bure Interferometer.
  IRAM is supported by INSU/CNRS (France), MPG (Germany) and IGN (Spain).
  This work has partially made use of 
  data products from the Two Micron All Sky Survey (2MASS), 
  which is a joint project of the Univ. of Massachusetts 
  and the Infrared Processing and Analysis Center/California Institute of 
  Technology, 
  funded by the National Aeronautics and Space Administration 
  and the National Science Foundation. 
  This research has also been partially supported by the Ministerio de
  Ciencia y Tecnolog\'{\i}a under the programmes AYA2004-01515 and 
  ESP2005-07714-C03-03 (including FEDER funds). 
  SMB acknowledges the support of the European Union 
  through a Marie Curie Intra-European Fellowship within the 
  Sixth Framework Program.

\end{acknowledgements}

\end{document}